# Tailored pore gradient in phenolic membranes for adjustable permselectivity by leveraging different poloxamers

Leiming Guo,[a,b,*] Martin Steinhart,[b] Yong Yang[c] and Liang Zhu[a]

[a] Key Laboratory of Integrated Regulation and Resource Development on Shallow Lake of Ministry of Education, College of Environment, Hohai University, Nanjing 210098, China

[b] Institut für Chemie neuer Materialien, Universität Osnabrück, Barbarastr. 7, 49069 Osnabrück, Germany

[c] Key Laboratory for Medicinal exploitation of Huai'an Resources, Faculty of Chemical Engineering, Huaiyin Institute of Technology, Huai'an 223003, Jiangsu, China

* Corresponding author

E-mail: leiming.guo@uni-osnabrueck.de

# Abstract

Cost-affordable phenolic membranes having gradient nanostructures can be facilely synthesized from resol oligomers in the presence of $ZnCl_2$ and poloxamers. The gradient nanostructures are formed by stacking phenolic nanoparticles with gradually enlarged diameters as the distance from the upper surface increases. The use of poloxamers for creating gelation surroundings is of great significance for controlling the growth of phenolic nanoparticles, which in turn dictates the performance of the phenolic membranes thus-produced. Hence, a study of the effects of poloxamers species on the preparation of the phenolic membranes is highly demanded since such robust membranes have much potential to be scale up for mass production. Herein, the poloxamer Pluronic F127 ($EO_{106}$-$PO_{70}$-$EO_{106}$; EO = ethyleneoxide, PO = propyleneoxide) was introduced in the membrane-forming formulations. As opposed to P123 ($EO_{20}$-$PO_{70}$-$EO_{20}$) that we used previously, F127 possessing extended PEO chains can delay the gelation during membrane formation. Hence, the phenolic nucleates are able to grow for longer durations, leading to the generation of more distinct gradient nanostructures in the phenolic membranes. Enhanced permeance can then be realized with F127-derived phenolic membranes. We also demonstrate that L31 ($EO_1$-$PO_{22}$-$EO_1$) with merely single terminal EO units at the ends of the PPO block could be used to prepare gradient phenolic membranes. This work is not only much helpful to deeply understand the design of the structural gradient in phenolic membranes, but capable of sheding light on the development of such intriguing structures for water purification.

# 1. Introduction

Membrane-based separation is one of the most important solutions in facing the urgent challenge of water scarcity worldwide [1]. In the separation process, membrane permeance and selectivity are of great significance and desirable to be drawn much attention [2-4]. Over the past decades, to realize high permeance as well as superior separation selectivity, composite designs have been identified as optimized membrane configurations, in which a selective layer and a porous support are combined [5]. Examples include self-assembled polymer membranes [6-8] silicon membranes [9,10], graphene oxide membranes [11-13], MXene membranes [14] and MOF membranes [15,16]. Most membranes with sufficient permselectivity suffer from high production costs. Drawbacks of composite designs are time-consuming procedures for their preparation and insufficient adhesion between the layers [17].

To tackle the aforementioned issues, asymmetric membranes containing integral gradient nanostructures have been developed [18]. It has been demonstrated that such membranes were able to be fabricated by time-saving and high-efficient methodologies, and were endowed with high permselectivity [19-21]. But one point should be noted that cost-affordable membrane materials still cannot be available for the production of these membranes, which, however, impedes the mass production of the asymmetric membranes with integral pore structures. Alternatively, as typically cheap and readily synthesized polymers, phenolics attract increasing attention and have been exploited to fabricate membranes for separation [22-26]. Due to the insoluble and infusible property of phenolics, it is hard to design gradient pore structures in phenolics. Thus the great limitation pertaining to the construction of the gradient structures in such cheap and easily available polymers should be addressed.

In our previous work, the gradient nanoporous phenolic membranes have been prepared by assembling resol oligomers, Pluronic P123 and $ZnCl_2$ [27]. During membrane-forming process, Pluronic P123 provided the gelation environment to control the nucleation of resol oligomers while $ZnCl_2$ was used to accelerate the thermopolymerization of resol oligomers and stabilize the system. Due to the faster growth of the nucleated phenolics in the membrane interior imposed by the concentration gradient of solutions in the course of ethanol evaporation, the phenolic nanoparticles with sizes gradually increasing with increasing distance from the upper membrane surface were then accessible. The gradient nanostructures were produced in the phenolic membranes by removing poloxamers and $ZnCl_2$ with acid soaking. P123 used for creating the gelation environment offered viscous surroundings, which prevented the excessive growth of the phenolic nanoparticles. To some extent, the gelation of P123 dictates the phenolic nanoparticle sizes and is, therefore, means to regulate membrane performances. Poloxamers with different lengths of PEO chains are supposed to exhibit varied gelation abilities [28]. Therefore, it is extremely intriguing to study the effects of the different poloxamers on membrane structures as well as permselectivity, which is also helpful in in-depth understanding of the mechanism of membrane formation as well as the extended applicability of the gradient nanoporous phenolics.

In this work, the poloxamer Pluronic F127 ($EO_{106}$-$PO_{70}$-$EO_{106}$) and L31 ($EO_1$-$PO_{22}$-$EO_1$) were employed to prepare the gradient nanoporous phenolic membranes. By adjusting the dosages of F127 and $ZnCl_2$ as well as the acid soaking durations, the F127-derived phenolic membranes with different pore gradients were prepared. Compared to the P123-derived membranes, the F127-derived ones exhibited enhanced gradient nanostructures and superior permeance, which is resulted from the delayed gelation of F127 during membrane formation. We further demonstrated that L31 with only one terminal EO group at each end of the PPO block still can cause the gelation of the membrane-forming system, resulting in the gradient nanoporous phenolic membranes. This work provides a pathway to strengthen the understanding of the formation mechanism of the gradient nanoporous phenolic membranes, and it may pave the way for the rational design of robust porous phenolic structures with tailored sizes for diverse applications.

## 2. Methods

**2.1 Materials.** $ZnCl_2$ ($\geq$ 98%) was acquired from Shanghai Xinbao Fine Chemical Factory. Ethanol ($\geq$ 99.8%) was purchased from Aladdin. Aqueous formaldehyde solutions (37 wt%) and NaOH ($\geq$ 96%) were obtained from Xilong Chemical Co., Ltd. Hydrochloric acid (36-37 wt%), phenol ($\geq$ 99%) and 98 wt% $H_2SO_4$ were purchased from Shanghai Lingfeng Chemical Agent Ltd. The poloxamer F127 and L31 were purchased from Sigma-Aldrich. Deionized water (conductivity: 8-20 μs/cm, Wahaha Co.) was used in all the experiments. Bovine serum albumin (BSA, molecular weight of 67 kDa, purity > 97%), cytochrome c (Cyt.c, molecular weight of 12.4 kDa, purity > 95%), and dextrans with four different molecular weights (10 kDa, 40 kDa, 70 kDa, and 500 kDa) were obtained from Sigma-Aldrich. Silicon wafers were cut into the size of 2 cm × 2 cm and then ultrasonicated at least three times in ethanol and dried before use.

**2.2 Preparation of ethanolic solutions of resol / $ZnCl_2$ / F127 and resol / $ZnCl_2$ / L31.** Resol was synthesized from phenol and formaldehyde by a base-catalyzed polymerization method, as described elsewhere [29,30]. In a typical synthesis process, 0.610 g (6.5 mmol) phenol was melted at 45 °C at first, and then 0.130 g of 20 wt% aqueous NaOH was added by stirring for 10 min. Subsequently, 1.054 g of 37 wt% formaldehyde was added to the above solutions dropwise. After stirring at 75 °C for 1 h, the reaction mixture was further cooled to room temperature followed by the pH adjustment with 0.6 M HCl toward pH 7. The mixture was dried under vacuum at 45 °C for 24 h to remove water. The as-prepared resol was redissolved with 10 g ethanol, which was then filtered with syringe filters with a nominal pore diameter of 0.22 μm to exclude out the insoluble NaCl precipitates from ethanolic solutions. After then, 1.1g ethanolic resol solutions were mixed with 2 g ethanol followed by addition of 0.08 g F127. At this time, the molar ratio of F127 to phenol was 0.01. Then 0.03, 0.05, 0.07, 0.08, 0.1 or 0.2 g $ZnCl_2$ were mixed with such solutions to produce the membrane-forming solutions. The molar ratios of $ZnCl_2$ to phenol were determined to be 0.34, 0.57, 0.77, 0.9, 1.1 and 2.2, respectively. To alter the dosages of F127, 0.08 g $ZnCl_2$ was added in 1.1 g diluted ethanolic resol solutions followed by addition of 0.03, 0.05, 0.08, 0.09 or 0.12 g F127, separately. And the molar ratio of $ZnCl_2$ to phenol is assigned as 0.004, 0.007, 0.01, 0.012 or 0.016, respectively. The corresponding compositions of each F127-containing membrane-forming formulation are summarized in Table 1.

**Table 1.** Compositions of F127-containing membrane-forming formulation used in this work

| Molar ratio of F127 to phenol | Molar ratio of $ZnCl_2$ to phenol | | | | | |
|---|---|---|---|---|---|---|
| | No.1 | No.2 | No.3 | No.4 | No.5 | No.6 |
| 0.01 | 0.34 | 0.57 | 0.77 | 0.9 | 1.1 | 2.2 |
| Molar ratio of $ZnCl_2$ to phenol | Molar ratio of F127 to phenol | | | | | |
| | No.7 | No.8 | No.9 | No.10 | No.11 | |
| 0.9 | 0.004 | 0.007 | 0.01 | 0.012 | 0.016 | |

To prepare the L31-derived membranes, the molar ratio of L31 to phenol was assigned as 0.095, the molar ratios of $ZnCl_2$ to phenol were adjusted to be 1.1, 1.6, 2.3, 2.8 and 3.4, respectively.

The corresponding compositions of each L31-containing membrane-forming formulation are summarized in Table 1.

**Table 2.** Compositions of L31-containing membrane-forming formulation used in this work

| molar ratio of L31 to phenol | molar ratio of $ZnCl_2$ to phenol | | | |
|---|---|---|---|---|
| | No.1 | No.2 | No.3 | No.4 |
| 0.095 | 1.1 | 1.6 | 2.3 | 2.8 |

**2.3 Fabrication of phenolic membranes.** To fabricate the phenolic membranes, 100 μL of the membrane-forming solutions were drop-cast on the cleaned Si substrates at room temperature and then instantly transferred into an oven preheated to 100 °C for the thermopolymerization of resol. After 12 h, the substrates together with the formed phenolic membranes were soaked in 55 wt% $H_2SO_4$ at 100 °C for 2 h to remove the poloxamer and $ZnCl_2$.

**2.4 Characterizations.** A field-emission scanning electron microscope (SEM, Hitachi S4800) was used to probe the surfaces and the cross sections of the samples at an accelerating voltage of 5 kV. Before SEM examinations the samples were sputter-coated with a thin layer Pd/Pt alloy to enhance the conductivity. Fourier transformation infrared (FTIR) spectra were obtained from a Nicolet 8700 infrared spectrometer in the attenuated total reflection (ATR) mode. The surface porosities of the varied phenolic membranes were estimated by binary images using *ImageJ*.

**2.5 Filtration and separation tests.** Permeability and rejection tests were performed on a Millipore filtration cell (Amicon 8003, Millipore Co.) at a stirring speed of 600 rpm and a pressure of 0.4 bar. A water storage tank was connected to the filtration cell for the continuous water supply during measurement. The filtration cell has a working volume of 3 mL and an effective membrane area of 0.9 cm$^2$. The phenolic membranes were attached on polyester non-woven supports before fitting into the filtration cell. A pre-compaction at 0.4 bar was carried out for 5 min to obtain a stable water flux, and then the permeances of the phenolic membrane were record. BSA and Cyt. C were dissolved in phosphate buffer (pH 7.4) at concentrations of 0.5 g/L and 0.02 g/L, respectively, and were used to probe the retentions of membranes prepared under different conditions. The BSA and Cyt.c concentrations in feeds and filtrates were monitored using a UV-vis absorption spectrometer (NanoDrop 2000c, Thermo) and the intensities of the BSA and Cyt.c characteristic peaks at the wavelength of 280 and 405 nm of the feeds and filtrates were compared to determine the retention rates of BSA and Cyt.c of each phenolic membrane.

Four dextrans with molecular weights of 10 kDa, 40 kDa, 70 kDa, and 500 kDa were mixed in water at a concentration of 2.5, 1.0, 1.0 and 2.0 g/L, respectively, and used to determine the molecular weight cut-off (MWCO) of the membranes. The concentrations of dextrans with different molecular weights were analyzed by gel permeation chromatography (GPC, Waters 1515).

## 3. Results and discussion

### 3.1 Gradient nanoporous phenolic membranes

Previously, we combined resol and $ZnCl_2$ with Pluronic P123 to design porous phenolic membranes with gradient nanostructures [27]. To reveal the impacts of poloxamers with different lengths of PEO chains on the structures and permselectivity of the phenolic membranes, in this work, we replaced P123 with F127 in the membrane-forming formulations. The molar ratios of F127 to resol and $ZnCl_2$ to resol will be systematically tuned. Considering resol has a wide distribution in molecular weight, phenol replaces resol to calculate the molar ratios (See experiment details).

When the molar ratios of F127 to phenol and ZnCl$_2$ to resol were initially designated as 0.01 and 0.9, respectively, the obtained phenolic membranes exhibited a rough surface composed of finely packed phenolic particles (< 10 nm) after soaking in H$_2$SO$_4$ (Figure 1a). The acid soaking was employed to remove F127 as well as ZnCl$_2$ in the membranes in such a way that the pores were ambiguously presented. On the membrane bottom surface, larger phenolic particles with a diameter of ≈143 nm were jointed to construct apparently porous structures (Figure 1b). The surface porosity of the membrane bottom was 44% as estimated from the binarized SEM images with *ImageJ*. The size discrepancy of the phenolic particles at the top and the bottom surfaces of the phenolic membrane indicates the presence of a structural gradient in the phenolic membrane. The structural gradient was further verified by inspection of the cross-sectional SEM images of the 10 μm thick phenolic membrane (Figure 1c and d), in which the sizes of the phenolic nanoparticles were gradually enlarged from the top to the bottom surfaces of the membrane.

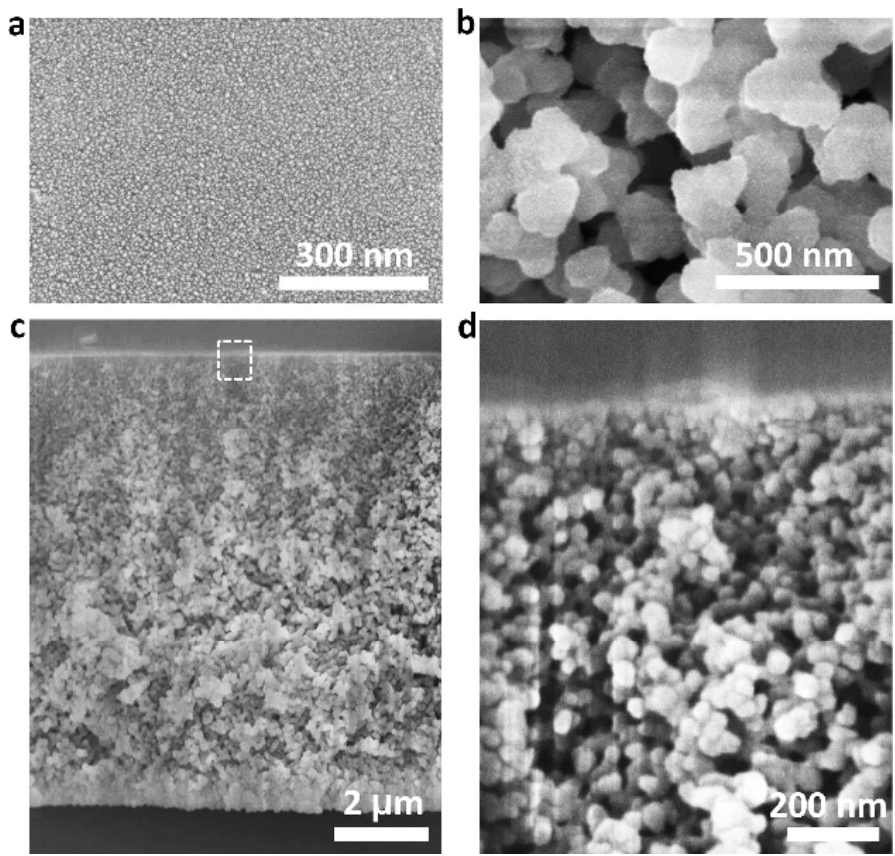

**Figure 1.** SEM images of the gradient nanoporous phenolic membranes prepared with molar ratios of F127: phenol = 0.01 and $ZnCl_2$: phenol = 0.9. (a) Top surface, (b) bottom surface and (c, d) cross sections of the membranes. (d) shows the area marked with dotted lines in (c).

Infrared spectroscopy (IR) was applied to monitor the changes of the chemical compositions of the phenolic membranes before and after acid soaking (Figure 2). The broad peak at around 3400 cm$^{-1}$ was strongly weakened after acid soaking, implying the removal of $ZnCl_2$ from the phenolic membranes by acid soaking. The peak at 1624 cm$^{-1}$ appeared due to the merging of the aromatic C=C band at 1605 cm$^{-1}$ and the characteristic –C=O band at 1648 cm$^{-1}$, which was induced by the aromatization and deoxygenation effect of $ZnCl_2$ on phenolics [31]. After acid soaking, however, the merged peaks recovered to their pristine positions [26]. Notably, the shifting of the peak from 1100 cm$^{-1}$ to 1079 cm$^{-1}$ occurred in the phenolic membranes before acid soaking, suggesting the complexation of Zn$^{2+}$ with F127 [26,33]. When soaking the phenolic membranes in the acid solutions, the peak at 1100 cm$^{-1}$ should be recovered, which, however, almost disappeared attributing to the removal of F127 from the phenolic membranes.

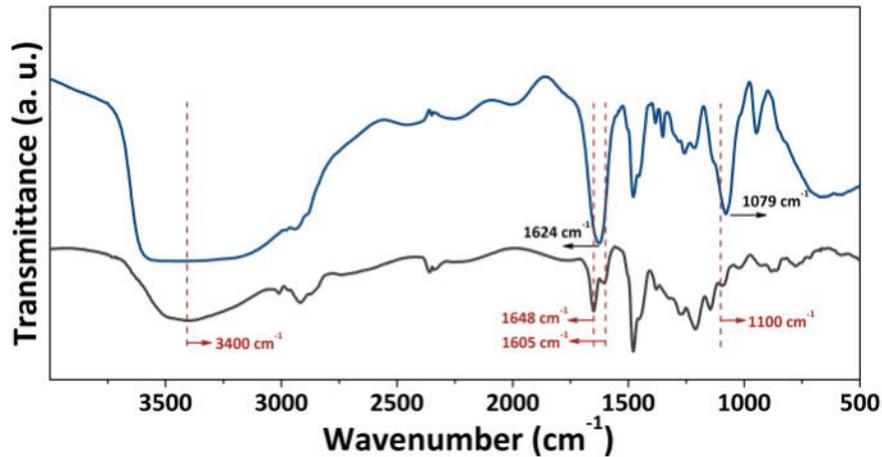

**Figure 2.** IR spectra of the phenolic membranes before (blue, top) and after (black, down) acid soaking.

**3.2 Evaluation of structural gradients of phenolic membranes**

To reveal the impacts of F127 and ZnCl$_2$ on the gradient structures of the phenolic membranes, the molar ratio of F127 to phenol was designated as 0.01, while the dosages of ZnCl$_2$ were tuned in the membrane-forming formulations. As the surfaces of the phenolic membranes always consisted of tiny nanoparticles with diameters < 10 nm, it is difficult to reliably distinguish the nanoparticle sizes on the top of each membrane by the microscopic methods that can be applied here. However, since the uppermost layers of the phenolic membranes were formed by fast F127 gelation and thermopolymerization [27], the nucleation and growth of phenolic nanoparticles at the top surfaces of the phenolic membranes likely occurs in a similar manner for all investigated configurations. The discrepancies of the separation selectivities of the different phenolic membranes likely arises from minute differences in the nanoparticles sizes at the top surfaces of the phenolic membranes. To evaluate the structural gradient of the phenolic membranes, we assumed that the average diameters of the phenolic particles at the top surfaces of the varied phenolic membranes ($d_{Top}$) are equal to each other, which will also be demonstrated by the following analysis. In this case, the membranes with enlarged phenolic particles at their bottoms will exhibit larger pore gradients. $G$ represents the pore gradient of the phenolic membranes, which is defined as below:

$$G = d_{Bottom} / d_{Top} \qquad (1)$$

Then the average diameters of the particles at the membrane bottoms ($d_{Bottom}$) are able to represent the structural gradient of the membranes. The relationship between G and $d_{Bottom}$ can be further described as follows:

$$G \propto d_{Bottom} \qquad (2)$$

This means the structural gradient $G$ is proportional to the average particle size at the membrane bottom. Therefore, $d_{Bottom}$ will be used as a measure to evaluate the structural gradient of the phenolic membranes in this study.

**3.3 Impact of ZnCl$_2$ dosages on structural gradients of phenolic membranes**

The bottom surface morphologies of the different membranes were then compared. With the increase of the molar ratios of ZnCl$_2$ to phenol, the diameters of the phenolic particles are gradually enlarged and then minimized (Figure 3a-e and Figure 1b). When the molar ratio of ZnCl$_2$ to phenol was 0.34, $d_{Bottom}$ was determined to be around ≈59 nm (Figure 3a). Upon increasing the molar ratios to 0.57 and 0.71, the particle sizes were further raised to ≈73 and ≈91 nm, respectively (Figure 3b and c), indicative of the enhanced gradient nanostructures in the membranes. As $ZnCl_2$ can accelerate the polymerization of resol and facilitate the nucleation of phenolic particles [27], the larger phenolic particles are formed and reside at the membrane bottoms. Meanwhile, the bottom surface porosities of 38, 37 and 35 % were acquired in the membranes at the molar ratios of $ZnCl_2$ to phenol = 0.34, 0.57 and 0.77, respectively (Figure 3f). The maximal porosity of 44% was found when the molar ratio came to 0.9. Further increasing the molar ratio of $ZnCl_2$ to phenol caused the reduction of the bottom surface porosity. For instance, the porosity was decreased to 37% when the molar ratio of $ZnCl_2$/phenol was assigned as 1.1. The reduced porosity is attributed to the excessive growth of the phenolic particles. The phenolic particles, however, were slightly increased from ≈143 to ≈147 nm in diameter when the molar ratio of $ZnCl_2$/phenol was raised from 0.9 to 1.1 (Figure 1b and 3d). The thermopolymerization of resol became faster and was continuously carried out during membrane formation when the molar ratio of $ZnCl_2$ to phenol was increased to 2.2, leading to a less pronounced porosity in the phenolic membranes (Figure 3e and Figure S1).

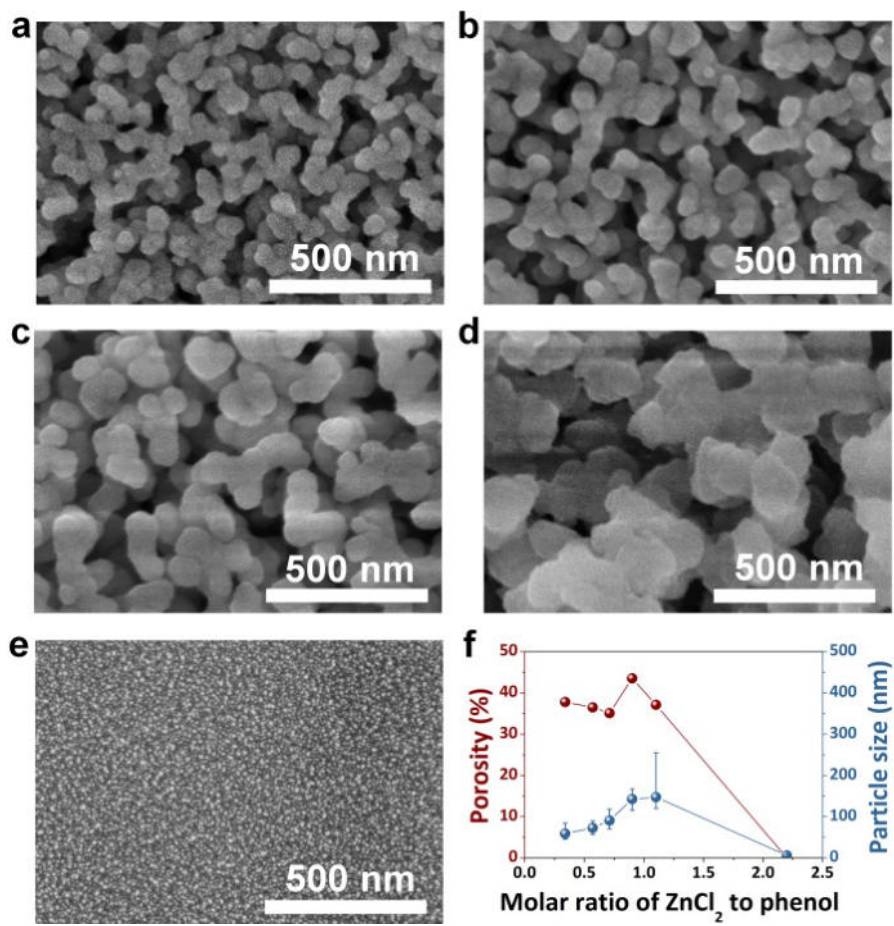

**Figure 3.** Phenolic membranes prepared with different $ZnCl_2$ dosages and at a fixed molar ratio of F127: phenol = 0.01. (a-e) SEM images of the membrane bottoms prepared at a molar ratio of $ZnCl_2$: phenol = 0.34 (a), 0.5 (b), 0.77 (c) 1.1 (d) and 2.2 (e), respectively. (f) Porosities and the particle sizes at the membrane bottoms as a function of the molar ratios of $ZnCl_2$ to phenol.

### 3.3 Impact of F127 dosages on structural gradients of phenolic membranes

To prepare the phenolic membranes by using different dosages of F127, a molar ratio of $ZnCl_2$ to phenol was designated as 0.9. Then the molar ratios of F127 to phenol will be systematically adjusted. With the increase of the molar ratios of F127 to phenol, the nanoparticle sizes of the prepared membranes went down initially and then leveled off followed by their further decrease (Figure 4a-d and Figure 1b). As the molar ratio of F127/phenol was 0.004, the insufficient gelation of F127 resulted in the fast nucleation of the phenolic particles. Consequently, the phenolic particles grew excessively, resulting in a surface porosity as low as 30% at the membrane bottom (Figure 4a and 4e). As the phenolic particles were merged with each other and then formed as a continuous network in the phenolic membrane, it is hard to determine diameters of individual phenolic particles. Alternatively, the diameters of the pore walls of the phenolic network can be considered as being representative of $d_{bottom}$. Thus, we measured the distances of the adjacent pores and obtained an average pore wall thickness of ≈173 nm. For the molar ratio of F127 to phenol was 0.007, $d_{bottom}$ decreased to ≈124 nm (Figure 4b), while the surface porosity was determined to be 29.4%. However, the increased dosages of F127 would inevitably induce the faster gelation and causes the reduction of the particle size [27]. For instance, the phenolic particle sizes were decreased to ≈85 and ≈35 nm at the bottoms of the membranes prepared at molar ratios of F127: phenol =0.012 and 0.016, respectively (Figure 4c and d). In stark constrast to the particle sizes, the bottom surface porosities of the two membranes went up to 43.5 % and 47.3 %, respectively.

It is to be expected that the PPO domains of F127 repel resol, whereas a certain enthalpic compatibility of resol and the PEO domains of F127 exits. However, mixing the PEO blocks with resol may require that the PEO blocks adapt entropically unfavorable stretched conformations; this effect likely reduces PEO/resol miscibility. Thus, phase separation between F127 and resol will occur, resulting in F127 phases consisting of PPO cores surrounded by PEO coronas in contact with the resol. This scenario is to some extent analogous to the micellation of amphiphilic block copolymers in selective solvents. In the case of low F127 dosages (the molar ration of F127: phenol = 0.004) discrete F127 entities surrounded by a continuous resol matrix form. After the degradation of the F127 by acid treatment, phenolic membranes are obtained that contain discrete pores formed in place of the volumes occupied by the discrete F127 entities (Figure 4a). If the F127 content is increased (the molar ration of F127: phenol ≥ 0.007), the F127 entities will touch each other and eventually form a continuous network. The F127/resol mixture has then a bicontinuos morphology reminiscent of early-stage bicontinuous morphologies formed by spinodal decomposition in symmetric binary mixtures. Degradation of the F127 by acid treatment then yields continuous pore systems (Figure 4b-d).

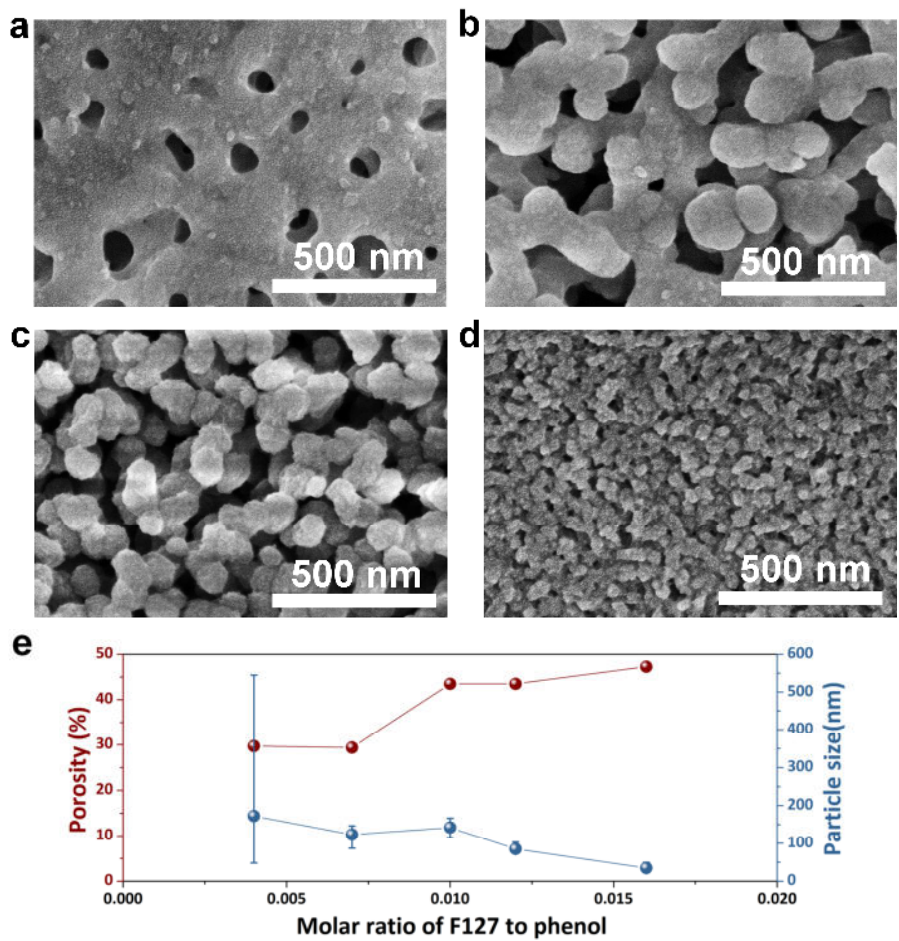

**Figure 4.** Phenolic membranes prepared with different F127 dosages and at a fixed molar ratio of $ZnCl_2$: phenol = 0.9. (a-d) SEM bottom surface images of the membranes prepared at a molar

ratio of F127: phenol = 0.004 (a), 0.007 (b), 0.012 (c) and 0.016 (d), respectively. (e) Porosities and particle sizes at the membrane bottoms as a function of the molar ratios of F127 to phenol.

**3.3 Impact of acid soaking durations on structural gradients of phenolic membranes**

In addition to the dosages of $ZnCl_2$ and F127, the acid soaking duration also influences the morphology and performance of the phenolic membranes. With the elongation of the soaking durations, $d_{bottom}$ firstly increased, followed by shrinking (Figure 5a-c and Figure 1b). For a duration of 1 h, a surface porosity of 31% and a $d_{bottom}$ value of ≈69 nm were found (Figure 5a). This insufficient soaking cannot remove F127 completely. By extending the soaking duration to 2 h, $d_{bottom}$ was dramatically increased to ≈143 nm, while a bottom surface porosity of 44% was acquired (Figure 1b). The increased particle size should be due to the foaming effect resulted from the treatment with concentrated sulfuric acid solutions. As the durations of acid soaking were further prolonged to 5 and 10 h, the particle sizes were reduced to ≈86 and ≈58 nm, respectively (Figure 5b and c). As the soaking duration is extended, possibly slow carbonization of the phenolic triggered by $H_2SO_4$ took place, giving rise to the shrinkage of the phenolic matrix [33]. Accordingly, the porosities of the bottom sides of the membranes were decreased to 38% and 37%. Notably, a similar structural evolution was also observed when Pluronic P123 was used for the preparation of phenolic membranes.

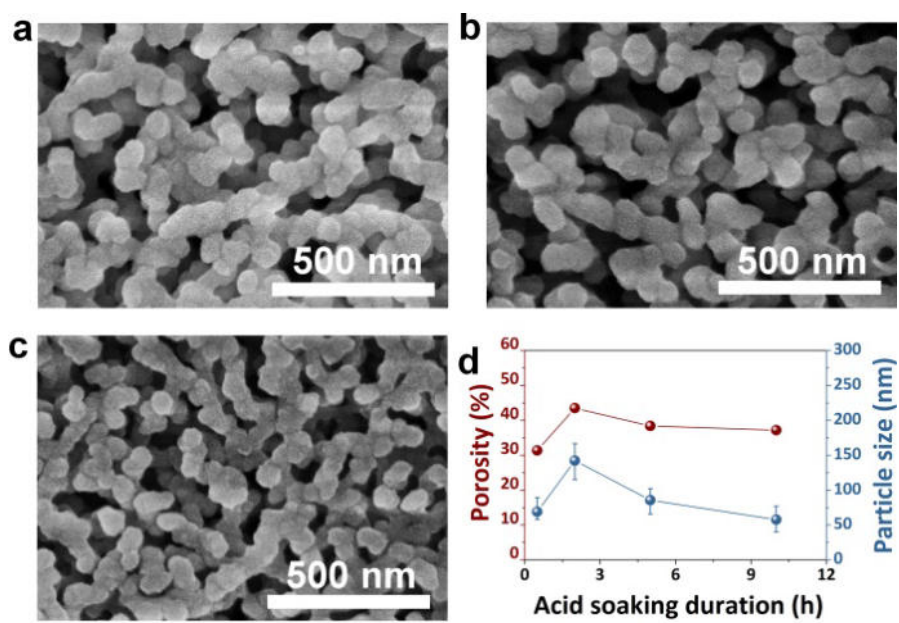

**Figure 5.** Morphologies of the bottom surfaces of the phenolic membranes prepared with molar ratios of F127: phenol= 0.01 and $ZnCl_2$: phenol = 0.9 after acid treatment with different durations. (a-c) SEM images of the membrane bottoms with the acid treatment of 1 h (a), 5 h (b) and 10 h (c), respectively. (d) Porosities and particle sizes at the membrane bottoms as a function of the durations of acid soaking.

### 3.4 Separation performances of phenolic membranes

The membrane performances were subsequently evaluated (Figure 6). With the increase of the molar ratios of $ZnCl_2$ to phenol from 0.34 to 0.57 and 0.77 while the molar ratio of F127 to phenol was fixed to 0.01, the water permeances were starkly enhanced from 625 L/(m²·h·bar) to 1152 and 1663 L/(m²·h·bar), respectively (Figure 6a). As $d_{bottom}$ of these membranes are gradually increased, the rising pore gradients resulted in the enhanced permeances. Moreover, the bovine serum albumin (BSA, $M_w$=66 kDa) rejections of these membranes amounted to 94, 99.5 and 91%, respectively. Upon the molar ratio of $ZnCl_2$ to phenol came to 0.9, the permeance as high as 1900 L/(m²·h·bar) was achieved due to the large structural gradient of such membranes. Meanwhile, the BSA rejection still retained as 95%. When the molar ratio of $ZnCl_2$ to phenol was continuously increased to 1.1, the contracted pore gradient caused a decreased permeance 1646 L/(m²·h·bar) as well as a reduced BSA rejection of 90%. Further increase of the molar ratio of $ZnCl_2$ to phenol to 2.2 resulted in a permeance as low as 96 L/(m²·h·bar), while the BSA rejection still amounted to 91%.

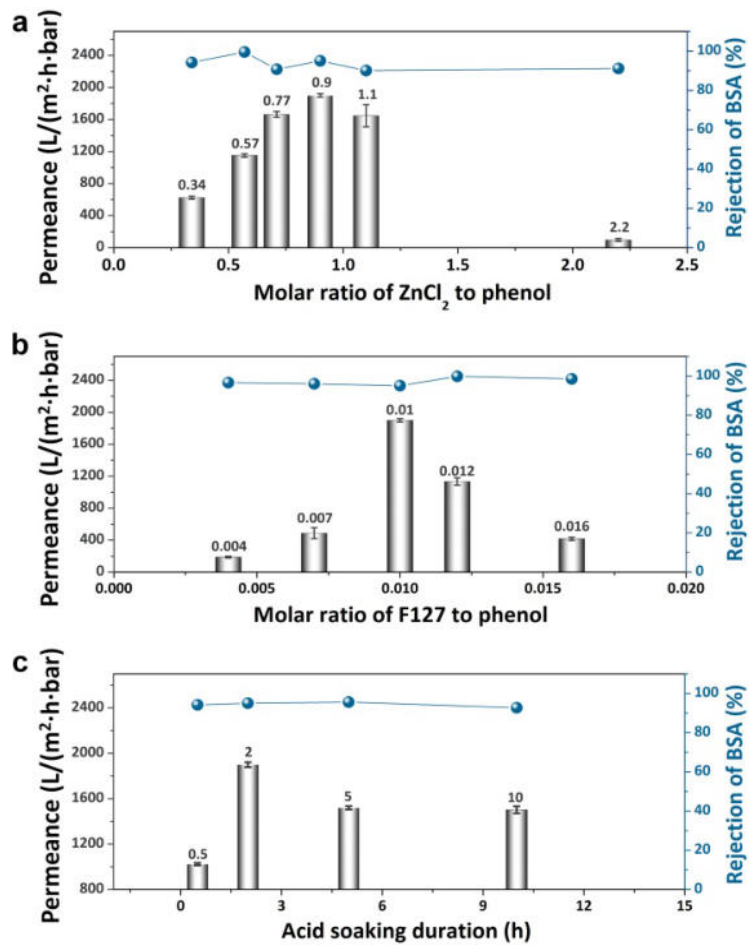

**Figure 6.** Performance of the phenolic membranes prepared under different conditions. (a) Water permeances and BSA rejections of the membranes prepared with different molar ratios of $ZnCl_2$ to phenol when the molar ratio of F127: phenol = 0.01. The acid soaking duration was 2 h. (b) Water permeances and BSA rejections of the membranes prepared with different molar ratios of F127 to phenol when the molar ratio of ZnCl$_2$: phenol = 0.9. The acid soaking duration was 2 h. (c) Water permeances and BSA rejections of the membranes prepared at the molar ratios of F127: phenol = 0.01 and ZnCl$_2$: phenol = 0.9 after acid soaking for different durations.

The water permeance of the thus-produced membranes exhibited a similar dependence on the molar with molar ratio of F127 to phenol as the values of $d_{bottom}$ (Figure 6b), implying the vital role of the pore gradient in determining water to pass through the membranes. With the increase of the molar ratios of F127 to phenol from 0.004 to 0.007 and 0.01, the permeances ranged from 188 to 487 and 1900 L/(m$^2$·h·bar), respectively. The BSA rejections of these membranes were further determined to be 96.6, 96 and 95%, respectively. Upon increasing the molar ratios of F127 to phenol to 0.012 and 0.016, the permeances of the membranes were decreased to 1134 and 417 L/(m$^2$·h·bar), respectively. And the BSA rejections can remain 99.8 and 98.5%, respectively.

It is noteworthy that poloxamers with shorter PEO chains possess stronger gelation capability [27]. As in the case of the PEO chains of P123 (20 repeat units per PEO block) were shortened compared to F127 (106 repeat units per PEO block), the gelation ability of F127 should be weaker than that of P123. If the molar ratios of F127 to phenol and P123 to phenol are fixed to 0.012 when preparing the F127- and P123-derived membranes, the $d_{bottom}$ exhibit a smaller value in the latter membranes. Thus the F127-derived membranes having larger structural gradients showed higher permeance than that of the P123-derived ones (707 L/(m$^2$·h·bar)). This indicates that the stronger gelation accelerates the termination of the growth of the phenolic particles during membrane formation, producing less pronounced structural gradients in the phenolic membranes. To further substantiate this standpoint, P123 was used to prepare the phenolic membrane at fixed molar ratios of ZnCl$_2$: phenol = 0.9 and P123: phenol = 0.01. The pore gradient of such membranes disappeared consequently after acid soaking (Figure S2). Under the same conditions, the membranes prepared by F127 have much more pronounced pore gradients. Hence, the weak gelation ability of F127 is a prerequisite for the generation of pronounced structural gradients in phenolic membranes.

As aforementioned, the phenolic membranes treated with 2 h acid soaking presented water permeance up to 1900 L/(m$^2$·h·bar). When the soaking duration was reduced to 1 h, F127 cannot be completely decomposed, the permeance of the phenolic membranes was thus decreased to 1024 L/(m$^2$·h·bar) (Figure 6c). However, the extended soaking durations, for example, 5 and 10 h, led to the shrinkage of the phenolic networks (Figure 5b and c), resulting in the decline of the water permeances to 1521 and 1502 L/(m$^2$·h·bar), respectively. Regardless of the varied soaking durations, the membranes exhibited the similar BSA rejections scattering around 95%. This outcome suggests that the morphology of the size-selective top surface of the phenolic membranes is hardly affected by the effects changing the morphologies of the phenolic membranes away from the top surface and $d_{\text{Bottom}}$ (Figure S3).

This notion is corroborated by the finding that the BSA rejections of all tested phenolic membranes are larger than 90%, indicative of the similar effective pore sizes in these membranes. Since only the permeances but not the selectivities are affected by $d_{\text{Bottom}}$, i.e., the structural gradient, the assumption underlying equation (2) is defensible, and permselectivities may be optimized by optimizing $d_{\text{Bottom}}$.

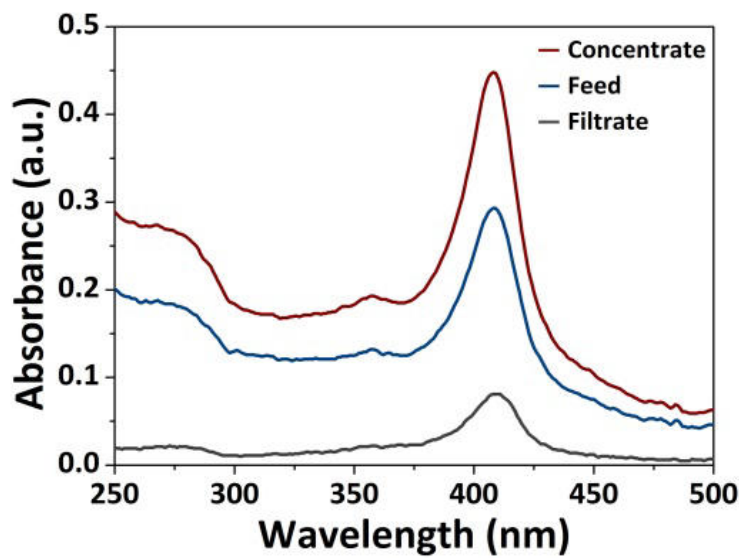

**Figure 7.** UV-vis spectra of Cyt.c in the concentrate, feed and filtrate.

To further assess the separation selectivity of the phenolic membranes, the membranes obtained from membrane-forming formulations with molar ratios of F127: phenol = 0.01 and $ZnCl_2$: phenol = 0.9 by acid soaking for 2 h were employed to concentrate cytochrome C (Cyt.c, $M_w$=12.4 kDa). Cyt.c has a lower molecular weight than BSA and passes through the membranes much easier. A rejection rate of 72.4% was obtained by comparing the intensities of the feed with the filtrate solutions (Figure 7). Dextrans with molecular weights of 10 kDa, 40 kDa, 70 kDa and 500 kDa were subsequently used as the rejection solutes to acquire the molecular weight cut-off (MWCO) of the membranes. As shown in Figure 8a, since the dextrans with higher molecular weights exhibit shorter responsive times, the response peak of the 500 kDa dextrans disappeared. As the response value is proportional to the concentration of dextrans, 500 kDa dextrans were completely rejected. Meanwhile, almost all of 40 and 70 kDa dextrans were rejected as their responsive values approached to 0. 10 kDa dextrans were also partly rejected on the basis of the response curve. According to the rejection curve (Figure 8b), the MWCO of such membranes was determined to be 31.2 kDa. Thus the effective pore diameter (D, Å) of the membrane can be estimated based on the equation as below [34]:

$$D = 0.3246 \times M^{0.557} \quad (3)$$

where $M$ (Da) represents MWCO. Then the effective pore size of such membranes can be roughly determined to be 10.3 nm, while the pore diameter of the P123-derived membranes with the highest permeance was estimated as 6.1 nm on the basis of the equation (3) [27]. Therefore, the rejection rates of BSA as well as Cyt.c of the F127-derived membranes are all slightly lower than those of the membranes fabricated with P123, but the permeance was much higher than that of P123-induced ones [27]. As a consequence, the F127-derived membranes are characterized by the larger structural gradients and the super permeances than P123-derived ones, while their selectivities are slightly reduced.

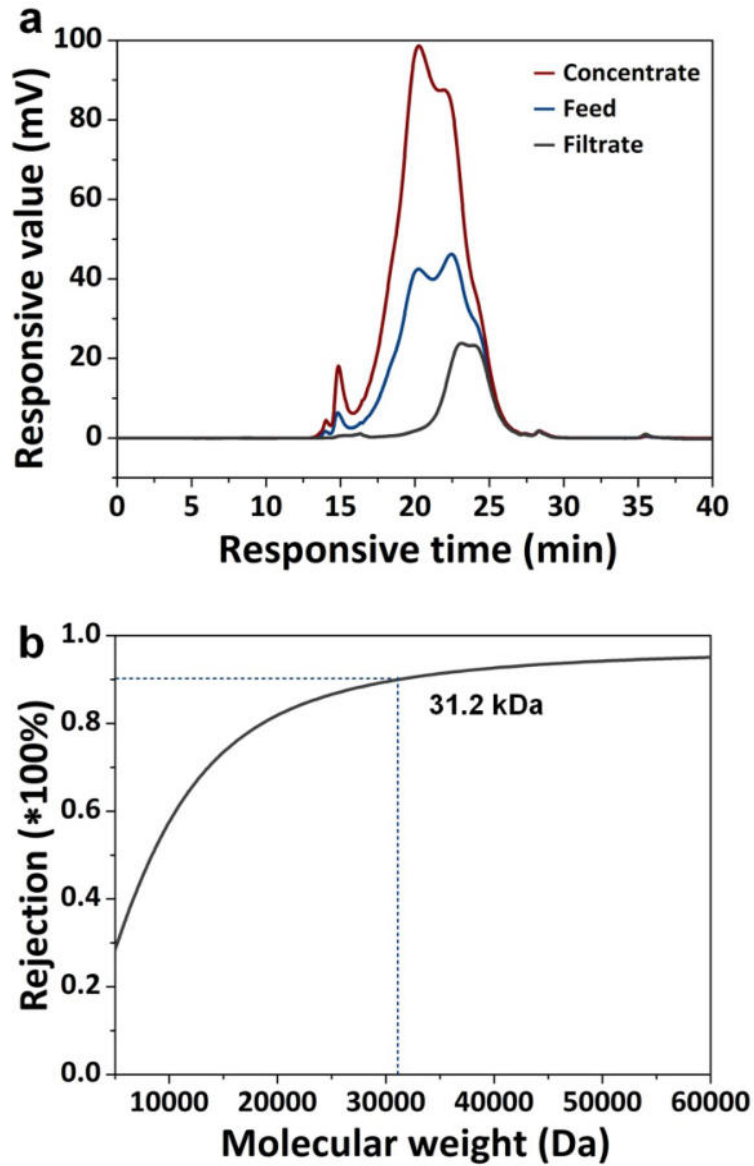

**Figure 8.** Rejection of dextran mixtures by the phenolic membranes. (a) Chromatograms of the dextran mixtures of the concentrate, feed and filtrate solutions. (b) Rejection curve of the dextran mixtures.

**3.5 Influence of different poloxamers on structures and separation performances of phenolic membranes**

To further reveal the effects of the PEO chains on the membrane structures, Pluronic L31 having two single terminal EO units ($EO_1$-$PO_{22}$-$EO_1$) was further used. We note that it is PEO rather than PPO that works in offering the gelation environment during membrane formation. We compared the membrane-forming formulations in which F127 or P123 was incorporated, and found that more $ZnCl_2$ and the poloxamers would be needed to prepare distinct structural gradients when the poloxamers with shorter PEO chains were employed. Here, more L31 should produce highly viscous surroundings during gelation to suppress the excessive nucleation of phenolic nanoparticles. The structural gradient can thus be yielded in the phenolic membranes. If L31 is insufficient, it cannot create a sufficient gelation environment to prevent the continuous nucleation of the phenolic particles since L31 is endowed with merely two EO units. Thus the molar ratio of L31 to phenol amounted to 0.095 initially. In this case, when the molar ratio of $ZnCl_2$ to phenol was 1.6, the gradient nanoporous structures are ambiguously presented in the membranes (Figure S4a). The porosity of the membrane bottom was determined to be 25%. An increased porosity of 41% was obtained when the molar ratio of $ZnCl_2$ to phenol was raised to 2.3 (Figure S4b). The distinct structural gradient was further demonstrated by the SEM cross section (Figure S4c). Upon increasing the molar ratios of $ZnCl_2$ to phenol to 2.8 and 3.4, the surface porosities of the membrane bottoms were assigned as 8 and 6%, respectively (Figure S4d and e).

Because of the stronger gelation capability of L31 as compared to that of F127 and P123, the nucleation of phenolic nanoparticles was terminated at an earlier stage during membrane formation, generating tiny phenolic nanoparticles in the membranes. As the molar ratios of $ZnCl_2$ to phenol were gradually increased from 1.6 to 2.3, $d_{bottom}$ were slightly enlarged to 27 and 29 nm, respectively. If the molar ratio of $ZnCl_2$ to phenol was continuously increased, the structural gradient gradually disappeared resulting from the acceleration of thermopolymerization induced by the excessive $ZnCl_2$. For instance, when the molar ratios of $ZnCl_2$ to phenol were 2.8 and 3.4, $d_{bottom}$ declined to ≈28 and ≈20 nm, respectively. This indicates that the pore gradients were weakened with the increase of $ZnCl_2$ dosages. The porosities and particle sizes of the L31-derived phenolic membranes are summarized in Figure S4f.

Notably, the most distinct structural gradient was obtained when the molar ratios of $ZnCl_2$ to phenol and L31 to phenol were assigned as 2.3 and 0.095, respectively. The water permeance of thus-prepared membranes amounted to 495 L/(m$^2$·h·bar) (Figure S4g), which is much lower as compared to the F127- or P123-derived membranes. The selectivity of the L31-derived membranes was then evaluated by the filtration of dextran mixtures. It is found that the MWCO of the membranes was determined to be 57.3 kDa when the highest permeance was obtained (Figure S5). The effective pore diameter of the L31-derived membranes was thus assigned as 14.5 nm according to the equation (3). We speculate that the excessively strong gelation ability of the poloxamer L31 prevents the fabrication of the membranes with enhanced separation selectivities. Higher L31 dosages yielded looser assemblyies of the phenolic nanoparticles, which likely enlarges the gaps between the nanoparticles. However, based on the aforementioned results, it is expected that poloxamers with suitable lengths of PEO chains may allow the preparation of the phenolic membranes with smaller effective pores toward nanofiltration (for instance, effective pore size < 2 nm).

## 4. Conclusions

Nanoporous phenolic membranes with structural gradients have been prepared by incorporating Pluronic F127 or L31 with $ZnCl_2$ and resol oligomers. The molar ratios of polxamers to phenol and $ZnCl_2$ to phenol in the membrane-forming formulations as well as the acid soaking duration were systematically evaluated to tailor the structural gradients and the perselectivities of the phenolic membranes. F127 with extended PEO chains has weaker gelation ability as compared to P123 used in the previous work, capable of producing the phenolic membranes with larger structural gradients. The F127-derived phenolic membranes having a thickness of 10 μm exhibited water permeance as high as 1900 (L/m$^2$·h·bar) as well as a MWCO of 31.2 kDa. Compared to the P123-derived membranes, the F127-derived ones presented higher water permeances at the little sacrifice of selectivity. When Pluronic L31 with shorter PEO chains and stronger gelation ability was employed to prepare phenolic membranes, the produced membranes were endowed with less pronounced structural gradients. As demonstrated in our work, the large-scale synthesis of gradient phenolic membranes can be realized by leveraging the cost-affordable ingredients. Therefore, the results reported here may pave the way for the design of the phenolic membranes with tailored structural gradients as well as adjusted permselectivities toward water purification.

## Acknowledgements

We acknowledge support by the European Research Council (ERC-CoG-2014, Project 646742 INCANA) and the Science and Technology Support Project of Jiangsu Provincial Sci. & Tech. Department (No.BY2016061-19).